\documentclass[review,12pt, 3p]{elsarticle}
\newcommand{\imwidth}{21pc}

\usepackage{amssymb}

\usepackage{graphicx}
\graphicspath{
	{images/}
}

\usepackage{subfiles}

\usepackage{upgreek}

\usepackage[hyphens]{url}

\usepackage{amsmath}
\usepackage{siunitx}

\usepackage{textcomp}
\usepackage{listings}
\journal{Digital Investigation}

\usepackage{subfig}
\begin{document}

\begin{frontmatter}



\title{Determining Image Sensor Temperature Using Dark Current\tnoteref{copyright}}


\author[eee]{R.~Matthews\corref{cor1}}
\ead{richard.matthews@adelaide.edu.au}
\author[eee]{M.~Sorell}
\author[cs]{N.~Falkner}
\cortext[cor1]{Corresponding author}
\tnotetext[copyright]{ Copyright 2018. This manuscript version is made available under the CC-BY-NC-ND 4.0 license http://creativecommons.org/licenses/by-nc-nd/4.0/}

\address[eee]{The University of Adelaide, School of Electrical and Electronic Engineering, Adelaide, SA, 5005 AUS.}
\address[cs]{The University of Adelaide, School of Computer Science, Adelaide, SA, 5005 AUS.}

\begin{abstract}
	The state of the art method for fingerprinting digital cameras focuses on the non-uniform output of an array of photodiodes due to the distinct construction of the PN junction when excited by photons. This photo-response non-uniformity (PRNU) noise has shown to be effective but ignores knowledge of image sensor output under equilibrium states without excitation (dark current). The dark current response (DSN) traditionally has been deemed unsuitable as a source of fingerprinting as it is unstable across multiple variables including exposure time and temperature. As such it is currently ignored even though studies have shown it to be a viable method similar to that of PRNU.  We hypothesise that DSN is not only a viable method for forensic identification but, through proper analysis of the thermal component, can lead to insights regarding the specific temperature at which an individual image under test was taken. We also show that digital filtering based on the discrete cosine transformation, rather than the state-of-the-art wavelet filtering, there is significant computational gain albeit with some performance degradation. This approach is beneficial for triage purposes.
	
\end{abstract}


\end{frontmatter}

\section{Introduction}

In this paper, we examine the relationship between temperature, dark current (DSN) and its effect on the accepted sensor pattern noise (SPN) methods using a discrete cosine transformation (DCT) filter instead of the computationally intensive wavelet filter. We achieve this using a lens cap applied during our image capture process to eliminate the interaction of light with our image sensor. We calculate the correlation across the temperature range of 10$^{\circ}$C to 50$^{\circ}$C in 5$^{\circ}$C increments before calculating a theoretical model for each of the cameras used in our experiments. This model is then used to contrast the photo-response non-uniformity (PRNU) SPN method against a DSN SPN only method similar to the hybrid SPN method in \cite{kurosawa2013casestudies}.


A reliable method of linking media to their source camera is through the analysis of pixel non-uniformity (PNU) sensor noise to generate a photo-response non-uniformity (PRNU) trace often referred to as a fingerprint \cite{lukas2006digital}. When tested across the limited range of -7.9$^{\circ}$C to 29.5$^{\circ}$C it has been observed that this method is not affected by temperature \cite{baabc}. \cite{lukas2006digital} goes as far as to state that PNU ``is not affected by ambient temperature or humidity'' due to the simple fact that PRNU is the dominant trace component of PNU. 

Sensor Pattern Noise (SPN) methods for solving the blind source camera identification problem has already been shown to be a valuable tool for both insurance providers and law enforcement. In \cite{dirik2009flatbed} it was shown that SPN for a scanned image differs to that of a genuine photograph. Such a method has applications for insurance fraud when detecting scanned images vs genuine images of goods; for example when attempting to prove ownership.  PRNU has previously been ruled out as a useful tool for policing insurance fraud concerning vehicle collisions due to the inability to link a camera to a particular vehicle \cite{mehrish2017multimedia}. In \cite{kurosawa2013casestudies} five separate cases are given which shows the practical application of SPN methods for law enforcement ranging from sexually based offences to those resulting in death.

Even though SPN is considered to be a robust and mature tool for linking images to cameras, has undergone significant peer review, has calculable potential error rates and has reached a level of general acceptance within the digital forensic community there are still questions regarding some features of the physics behind the method. \cite{kurosawa2013casestudies} expanded upon the method above to generate a PNU hybrid trace model based on the inhomogeneous nature of both PRNU and Dark Current (DSN). In \cite{matthews2018analysis} it was demonstrated that DSN exhibits signal power which adds to the overall correlation energy during the original SPN methods even when using sensors that have DSN correction methods. Since it is accepted that DSN is heavily temperature dependent, we investigate whether the current SPN methods are immune to temperature bias. 

The rest of the paper is organised as follows. In Section \ref{RelatedWork} we document the prior efforts in this field and highlight our novel contribution. In Section \ref{materials} our methods and experimental set-up is documented.   The results of our experiments are presented in Section \ref{results} before being discussed in \ref{discussion}. Finally, future work is presented, and the paper is concluded in Section \ref{conclusion}.

\section{Related Work}\label{RelatedWork}

\begin{figure}[!ht] 
	\centering
	\subfloat[][]{	
		\includegraphics[width=7.5cm]{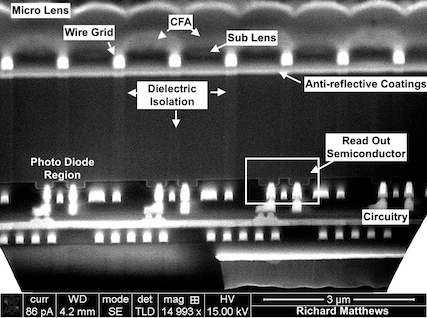}
	}
	\qquad
	\subfloat[][]{	
		\includegraphics[width=7.5cm]{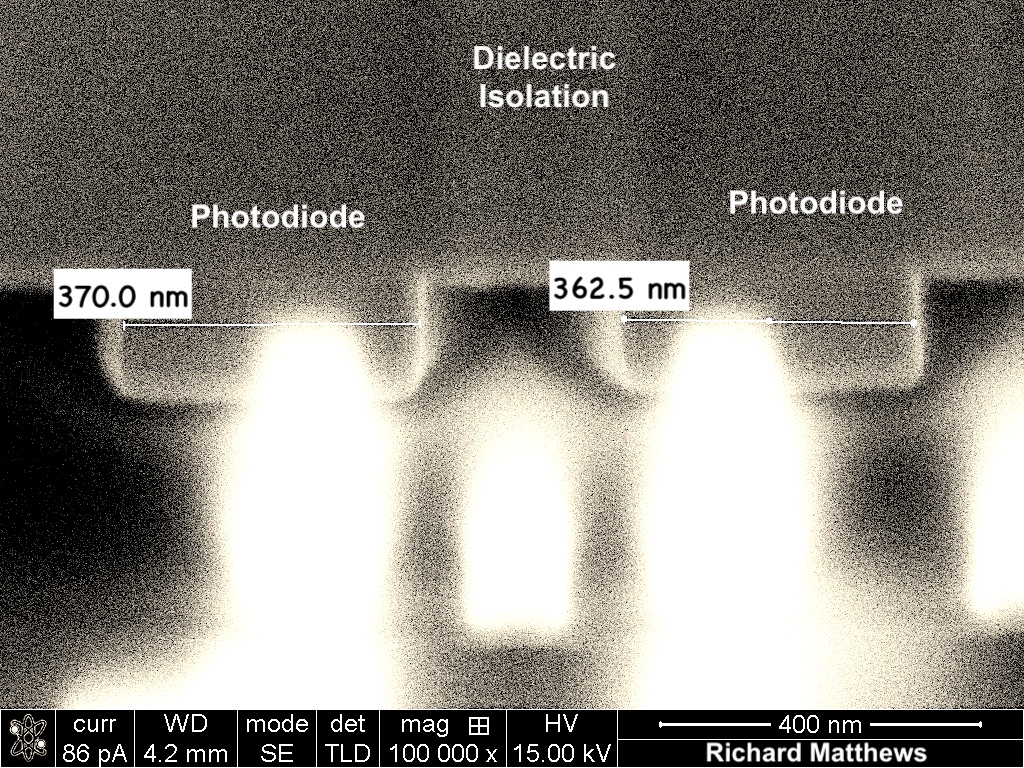}
	}
	\caption[]{(a) A cross section of a CMOS image sensor. (b) Inconsistencies within the diode region can be visually observed between neighbouring pixels. While this will not effect the overall function of the sensor this is an example of pixel non-uniformity (PNU).}
	\label{Microsopicimage}
\end{figure}

A reliable method of linking images to their source camera is through the analysis of pixel non-uniformity (PNU) sensor noise to generate a PRNU trace often referred to as a fingerprint \cite{lukas2006digital, kurosawa2013casestudies}. The basic premise of these methods is to apply a high pass filter to an image which isolates a noise residue directly related to the non-uniform nature of how the sensor outputs an electrical signal. This non-uniform output is due to manufacturing defects such as misalignment and inconsistent silicon dopant across a wafer during production \cite{matthews2018rethinking}. An example of this can be seen in Figure \ref{Microsopicimage}.

A photo-diode consists of a junction of positively and negative charged semiconductor material to form a depletion region. When a photon enters this depletion region it produces an electron-hole pair by transferring the energy of the photon to an electron resulting in the electron moving to a higher valence band or even becoming a free electron. We refer to these free electrons generated from photon energy as $e_{PH}^{-}$.  $e_{PH}^{-}$ is stored in the N-well semiconductor region of the photodiode which causes the depletion region to shorten.The associated hole left by $e_{PH}^{-}$ moves to ground via the P-type semiconductor. We refer to electrons generated due to the swapping of minority carriers without external excitation as dark electrons $e_{DARK}^{-}$  the movement of which promotes dark current. The n-well region is gradually filled to capacity $N_{max}$ by the combination of these electrons at which point the depletion region is removed:

\begin{equation}
N_{max}= N_{e_{PH}^{-}}+ N_{e_{DARK}^{-}}
\end{equation}

This combination of $e_{PH}^{-}$ and $e_{DARK}^{-}$ results in a sensor output PNU that is linked both to the sensor's PRNU and DSN. We have discussed how DSN is included in the SPN noise model in our previous work \cite{matthews2018rethinking}. In our previous work \cite{Matthews2017Isolation, Matthews2018Isolation}, we have also shown that the state of the art method for generating noise residue traces based on PRNU \cite{goljan2016effect} contains additional forensic information in the form of lens aberrations and dark current even when dark current removal or suppression techniques are employed on the integrated circuit (IC). This result verified the work previously highlighted in \cite{kurosawa2013casestudies, knight2009analysis} where it was shown that additional information could be obtained via a hybrid SPN DSN method to solve the blind source camera identity problem in a real-world setting.

To isolate the PNU effects, a three-stage process is used. The first stage is applying some form of filter to obtain a noise residue via the formula:

\begin{equation}
Y = I - f(I)
\end{equation}

where $Y$ is the noise residue obtained containing the SPN signal and $f()$ is the filter used to isolate the noise in the image. The second is the estimation stage where the SPN is estimated from a set of noise residues to remove the effects of random variables. Finally, the third stage is the post-estimation phase where the SPN can be enhanced for more accurate and precise camera identification. 

Focusing on the first stage of this process much work has been done in the area of signal processing to provide alternative filters than the original filter based on a wavelet corring method seen in \cite{lukas2006digital}. The work of \cite{li2010source} demonstrated the need for accurate highpass filtering since an image can contaminate the estimated PRNU if it contains significant high spatial frequency content such as edges, lines, contours and texture. Similarly, in the work of \cite{Matthews2017Isolation, Matthews2018Isolation} edge effects of the image are taken into account before filtering to ensure effects such as ringing are taken into account within the estimate. A well-written summary of the state of the art regarding different filter techniques is discussed in the background work of \cite{lawgaly2017sensor} before proposing an improved locally adaptive DCT (LADCT) filter and documenting its effectiveness. 

This noise residue is also susceptible to high-frequency patterns such as those generated through JPEG compression \cite{knight2009analysis, khanna1740forensic}. JPEG compression, being a lossy compression algorithm, will remove high-frequency components and thus lowers a potential correlation match between source and reference. All of this additional information is of value to a forensic investigation as it enables a more confident match to be established between camera and image in the context of the blind source camera identification problem when correctly accounted for. Understanding how to account for these additional sources of potential bias, however, is left to a suitably trained investigator and reinforces the already established work as seen in \cite{lukas2006digital, khanna1740forensic, alles2009source}.

When tested across the limited range of -7.9$^{\circ}$C to 29.5$^{\circ}$C it has been observed that this method is not affected by temperature \cite{baabc}. However, the theory as shown in \cite{holst2007cmos} indicates that DSN is not immune to temperature and in fact bears an exponential relationship due to the relationship between dark current density and temperature following the equation:

\begin{equation}\label{DarkCurrentDensity}
J_{D} \propto T^{2}e^{\frac{(E_{t}-E_{G})}{kT}}
\end{equation}

The Scientific Working Group on Digital Evidence (SWGDE) has previously released an error mitigation framework \cite{SWGDE_ErrorMit}, which introduces a basic strategy to identify and mitigate sources of likely error within digital forensic tools. Among the quality control and tool testing measures described is finding ``untested scenarios that introduce uncertainty in tool results.'' In this paper, we conduct tool testing of the SPN method in high-temperature environments to determine if the method is immune, particularly to the variances in temperature seen in vehicles during an Australian Summer. In this scenario, the dashboard of a car is known to fluctuate from 19$^{\circ}$C due to climate control, to more than 60$^{\circ}$C. This is known from the practical experience of the first author working in a proprietary setting for industry. 


\section{Research Methodology}\label{materials}

\begin{figure}[!ht] 
\centering
\includegraphics[width=\imwidth]{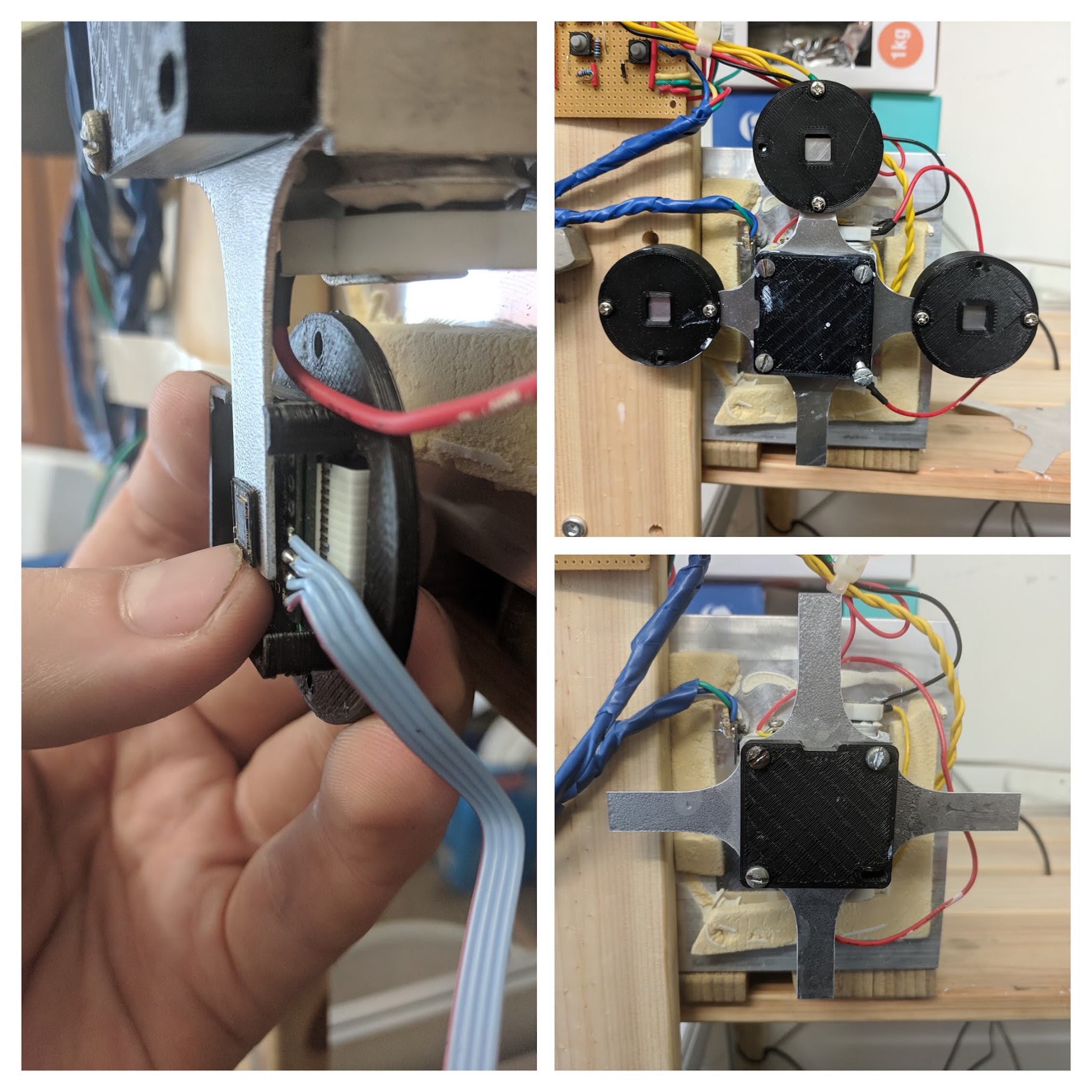}
\caption[Gaussian High Pass Filter in DCT Domain]{The Arduino controlled Peltier plate device used for controlling the temperature of the cameras. Seen here is the mounting position used for the image sensor to obtain an accurate reading of the sensor as opposed to the PCB. The black mounting square (shown here) was replaced with an aluminium block for thermal sinking purposes before imaging was conducted.}
\label{Arduino}
\end{figure}

To obtain a dark current signal which is uncoupled from other signals a series of steps are followed based on the image pipeline model of a digital camera. We use three Sony IMX219 CMOS digital image sensors mounted onto a Peltier plate temperature controlled device. The image sensors have built-in low dark current by design \cite{sonyIMX219} through the use of correlated double sampling both before and after the Analogue to Digital Converter \cite{EXMOR}.  Through the use of an Arduino controlled Peltier plate device we vary temperature between 10$^{\circ}$C and 50$^{\circ}$C in 5$^{\circ}$C increments. This Peltier plate is attached to a metal plate which extends into thin fingers of metal that the IC of the image sensor is secured on using a custom 3D printed enclosure and thermal paste. The IC itself is located onto the metal finger as opposed to the PCB of the camera to ensure the temperature of the sensor is captured as opposed to the temperature of the PCB the sensor is mounted upon. To measure the temperature of the sensor an MCP9808 solid state temperature sensor is mounted on the reverse side of the metal finger underneath the sensor IC. This setup is shown in Figure \ref{Arduino}.

The aperture of the camera is covered with several layers of black electrical tape to ensure no photons are allowed to enter the imaging column. Covering the aperture ensures only dark frames are captured. Using a python script, we set exposure time to a constant t = 1/1008s, and the effects of internal amplifiers are controlled by setting a predetermined long wait time during the setting up of the camera to allow the gains to reach a stabilised point before setting the ISO light sensitivity to 800. Future implementations of this experiment should take advantage of the additional code to be included in a future release of the Raspberry Pi camera distribution \cite{RasPi} to allow the manual setting of the Analogue and Digital gains \cite{pigain}. Each image is saved as a JPEG with 100QF setting with appended BAYER raw information to the end of the JPEG file. JPEG is used due to a limitation of the Raspberry Pi Camera API. While it is noted that JPEG at QF100 is not the same as lossless JPEG, the appended BAYER raw information is extracted to obtain a RAW format image using DCRAW \cite{DCRAW}. This BAYER raw format thus avoids all onboard processing steps of the digital pipeline within the image pipeline model of the Raspberry Pi Camera model \cite{Matthews2018Isolation}. To extract the RAW information from JPEG file it is converted to TIFF using DCRAW as per \cite{knight2009analysis} using the command:

\begin{center}
	\textbf{\textit{dcraw -D -T -4 -W filename}}
\end{center}

\begin{figure}[!ht] 
\centering
\includegraphics[width=\imwidth]{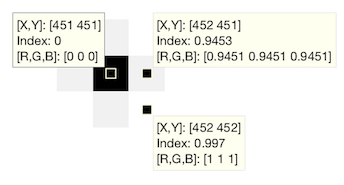}
\caption[Gaussian High Pass Filter in DCT Domain]{The Gaussian high pass filter with cut off frequency (150/1136)pi in the DCT domain. The cut off frequency is chosen to replicate that of the wavelet corring method of \cite{lukas2006digital}.}
\label{Guassian}
\end{figure}

For each temperature interval, an image set of 100 dark frames is taken per camera. We prepare a noise residue for each dark frame by filtering each image using a high-pass filter in the discrete cosine domain to extract the high-frequency components using Matlab. This filter is similar to the one seen in \cite{lawgaly2017sensor} however we only employ a simple image mask as demonstrated in Figure \ref{Guassian}. A cut off frequency of 150/1136pi is chosen to match the seminal work of \cite{lukas2006digital}. This mask is applied purely in the DCT domain as an implementation of ``goldilocks'' filtering. The hypothesis is that simple filters run in quicker time with fewer resources required and thus can serve as a useful tool for processing of large evidence datasets.

The effectiveness of the DCT filter is demonstrated in figure \ref{AccuracyPrecision}. Pixels that are greater than 95\% full value are considered to be saturated and are ignored as hot pixels \cite{kurosawa1999ccd, kurosawa2013casestudies}. Excluding saturated pixels ensures we are matching the unique DSN as opposed to the current literature of saturated pixels. The DSN noise residues are then averaged to obtain a single reference pattern for each temperature interval. It is noted that 100 dark frames may be excessive and a smaller number is potentially viable, however, this is excluded from this study.

We then correlate the reference pattern against a different set of illuminated flat field images captured at 30$^{\circ}$C and the same exposure time as the dark frames. The flat field images were taken using six discrete interchangeable lenses per camera resulting in a total of 300 images per camera, 50 per lens, per camera. A total of 2,700 correlations were made across the three cameras between the lens images and the dark current reference patterns. It is observed that the lens affects the data (Figures \ref{MultiLens}); to offset this effect the raw data across all lenses are treated as a single data set for each camera. This raw data is presented as box plots in Figures \ref{AllCameraBox}. The result of these correlations is then averaged for each camera to obtain a single result per temperature interval. These results are shown in figures \ref{Camera01DSN}, \ref{Camera05DSN} and \ref{Camera06DSN}.

\section{Data Collection and Analysis}\label{results}

\begin{figure}[!ht] 
\centering
\includegraphics[width=32pc]{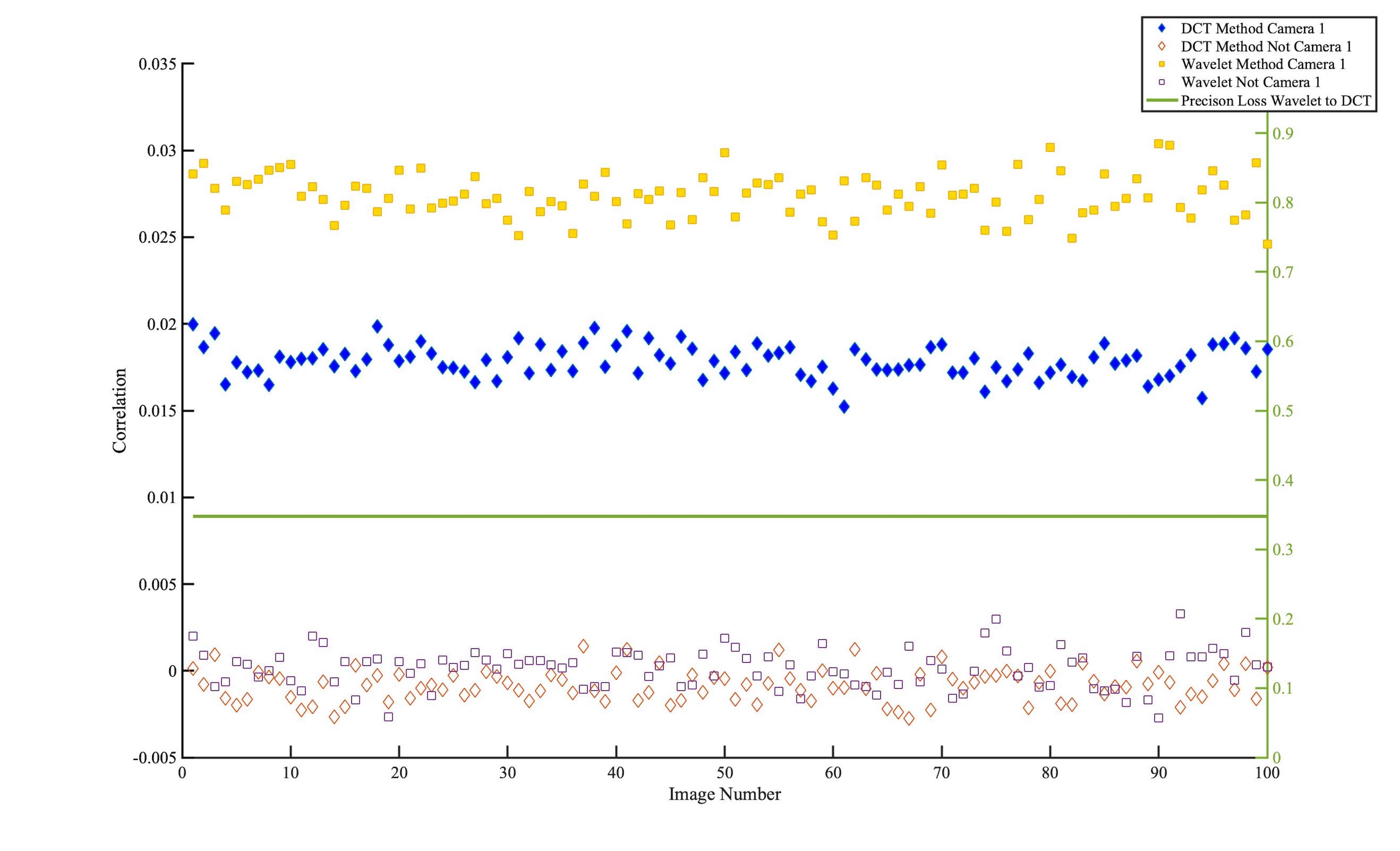}
\caption[Comparision of Wavelt and DCT method]{A typical comparative example of the accepted wavelet filter as shown in \cite{lukas2006digital} compared to our DCT Gaussian filter.}
\label{AccuracyPrecision}
\end{figure}

The results of the comparative test between the wavelet coring method and the proposed DCT method used in the rest of this paper is displayed in Figure \ref{AccuracyPrecision}.

A comparison of how long each method took to execute is displayed in Table \ref{runtimes}. This table when read in conjunction with Figure \ref{AccuracyPrecision} shows a computational gain with a loss of correlation. Such a trade off is of benefit for large data sets.

\begin{table}[!ht] 
\centering
\caption{Run Times of Filter Methods} 
\begin{tabular}{cccc} 
	\hline
	Filter Run & Wavelet (s) & DCT (s) & Delta (s)\\ 
	\hline
	Set 01     & 630 & 272& -358 (-57\%)  \\
	Set 02  & 832 & 315   & -517 (-62\%)\\
	\hline
\end{tabular}
\label{runtimes}
\end{table}

Since a large-scale test of similar filters has already been conducted in \cite{lawgaly2017sensor} we refer to those results as to the suitability of this method. The results shown here are illustrated merely as confirmation that the tool, as built, is capable of performing the experiment as designed in the remainder of the paper. We leave a complete, robust test of this ``goldilocks'' filtering method as future work.

The following observations are made. The DCT filter used may present itself as a viable method for triage since the DCT filter on average took between 4 and 6 minutes to execute whereas the wavelet method took between 10 and 14 minutes over the same image data and same resource assignment. The DCT filter is less precise, however, the accuracy of the method is not affected. The variability in time is due to local resources assigned from the High-Performance Computer cluster beyond the control of the operator. The difference in execution times indicates a speedup of approximately 60\% for a loss of 37\% precision. Such numerical values are quantified only for this specific hardware and need to be considered across any resource being allocated. 

\begin{figure}[!ht] 
\centering
\includegraphics[width=42pc]{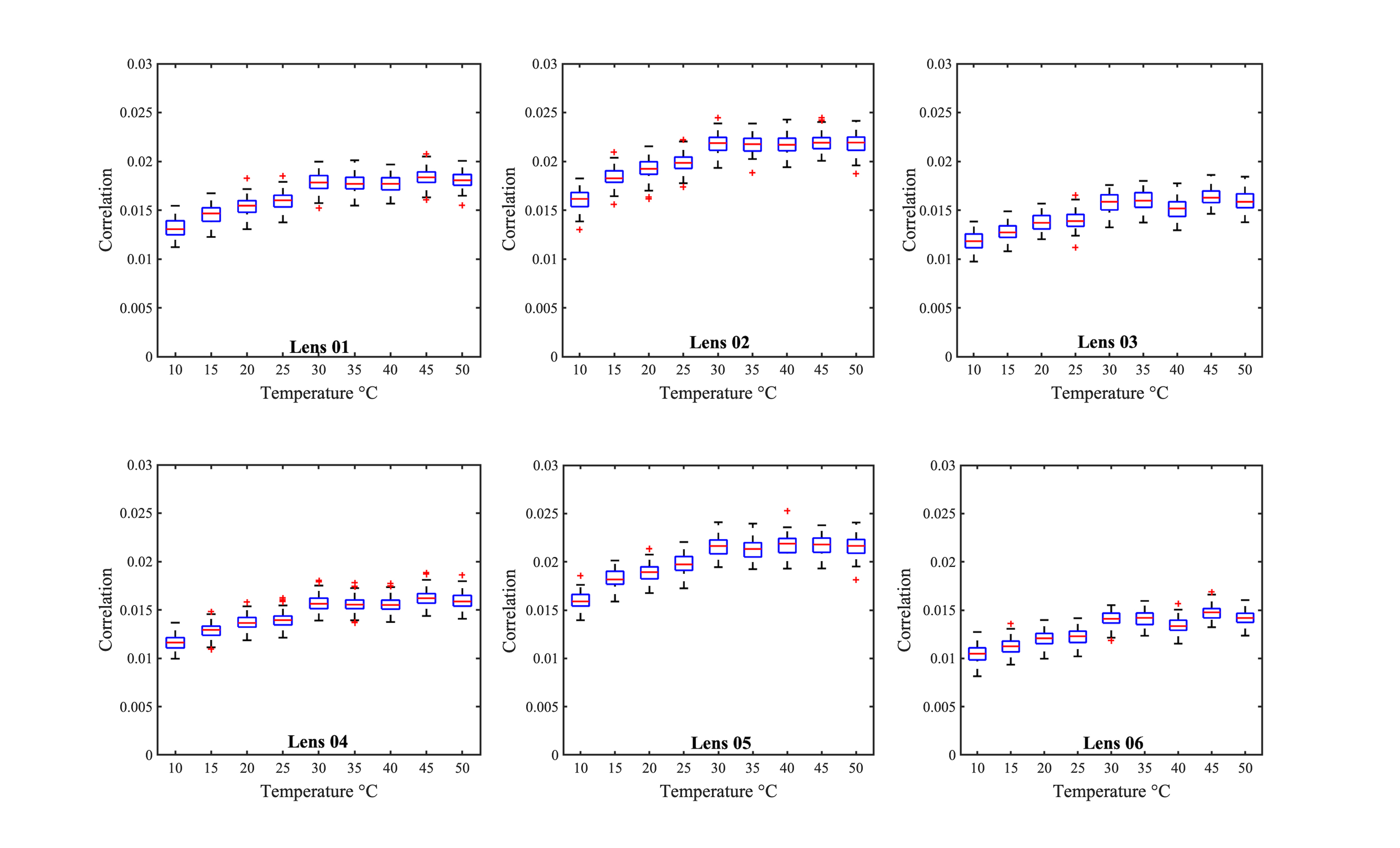}
\caption[]{Correlation verses temperature plot for the various lenses across camera one. It is clearly evident that even with only a DSN reference pattern that the lens still plays a role in unique identification reinforcing the results of \cite{Matthews2017Isolation,Matthews2018Isolation, matthews2018rethinking}.}
\label{MultiLens}
\end{figure}

The results of a single camera using the six different lenses are shown in Figure \ref{MultiLens}. This graph omits the correlations from the non matched camera for clarity. These omitted results are zero mean around or just below the zero axes. 

\begin{figure}[!ht] 
\centering
\includegraphics[width=42pc]{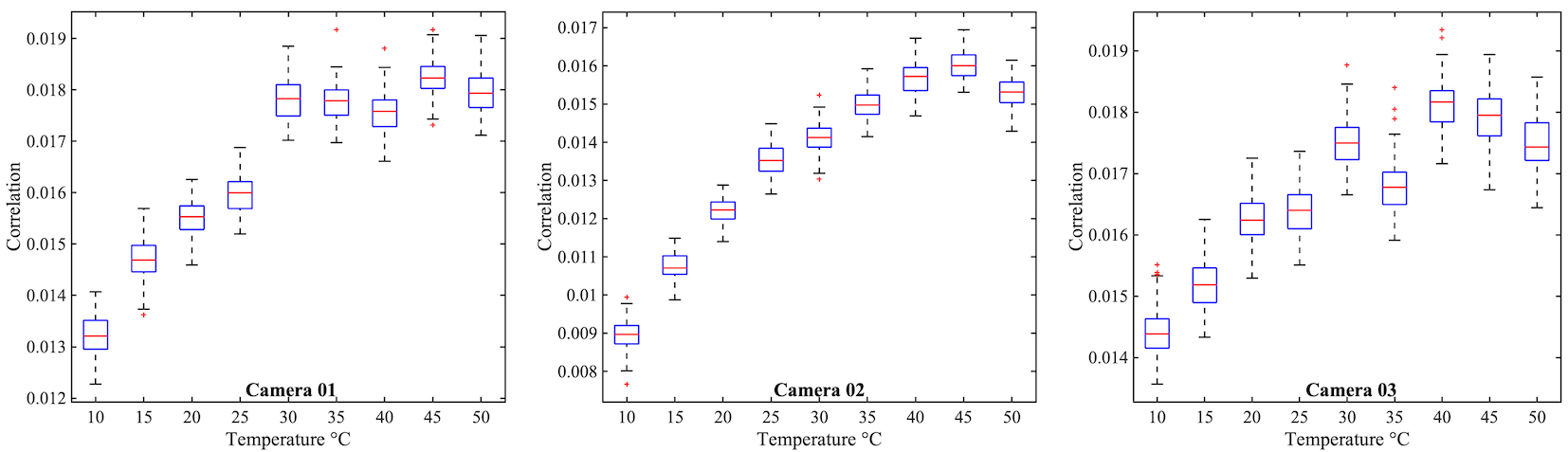}
\caption[]{Correlation verses temperature plot across the three cameras.}
\label{AllCameraBox}
\end{figure}

The results of each camera used in our method as a blind identified camera to a dark current reference pattern are shown in Figure \ref{AllCameraBox}. This graph omits the correlations from the non matched cameras for clarity. These omitted results are zero mean. 

\begin{figure}[!ht] 
\centering
\includegraphics[width=32pc]{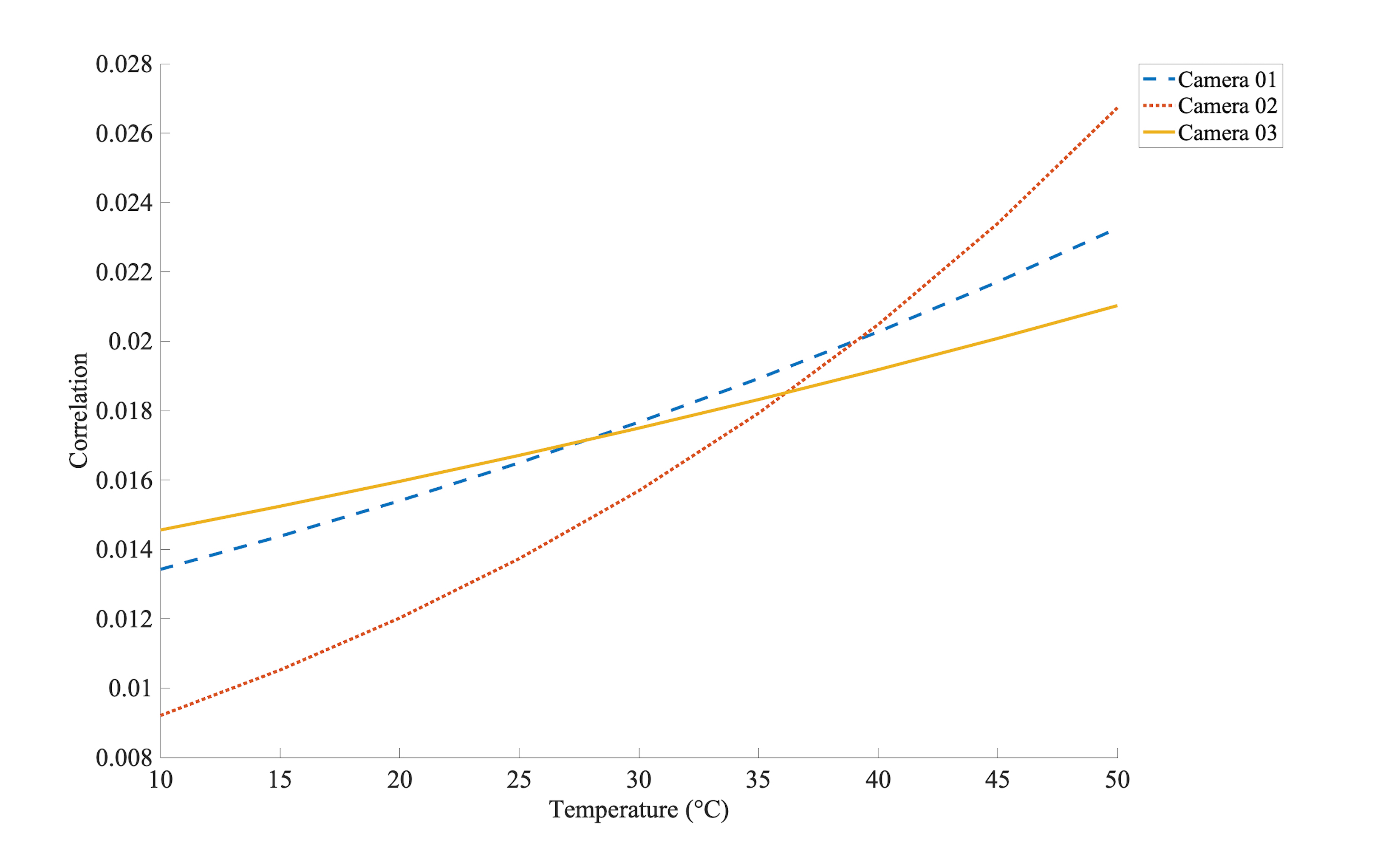}
\caption[]{The three theoretical curves plotted against each other enabling an indication of the dopant strength to be determined.}
\label{AllCamerasDSN}
\end{figure}

\begin{figure}[!ht] 
\centering
\includegraphics[width=32pc]{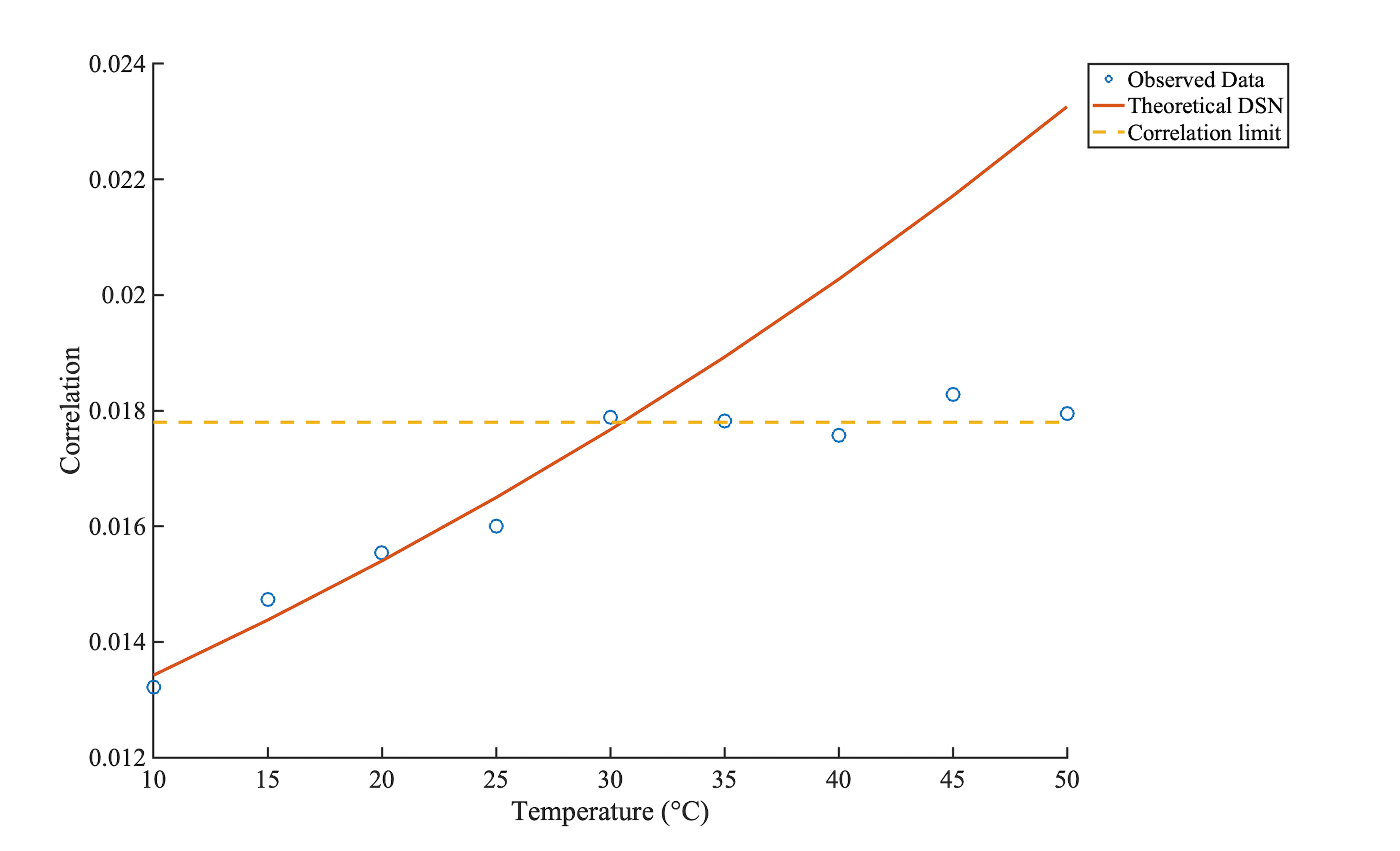}
\caption[]{Correlation verses temperature plot for camera one showing the correlation increase in accordance with the theoretical model to a limit which corresponds to the temperature of the image sets under test. Camera 1 Identified Temperature ~30.5$^{\circ}$C}
\label{Camera01DSN}
\end{figure}

\begin{figure}[!ht] 
\centering
\includegraphics[width=32pc]{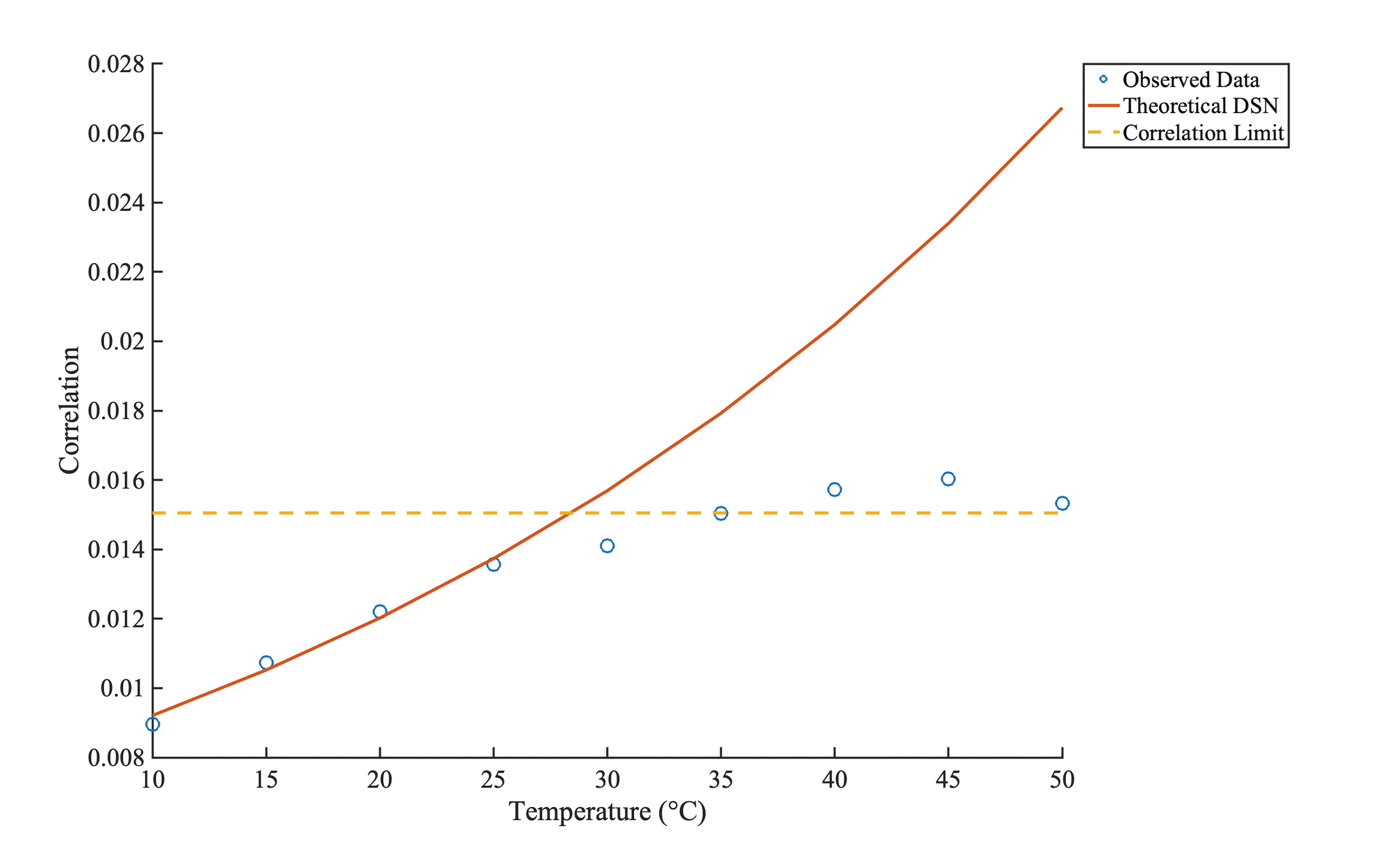}
\caption[]{Correlation verses temperature plot for camera two showing the correlation increase in accordance with the theoretical model to a limit which corresponds to the temperature of the image sets under test. Camera 2 Identified temperature ~28.35$^{\circ}$C}
\label{Camera05DSN}
\end{figure}

\begin{figure}[!ht] 
\centering
\includegraphics[width=32pc]{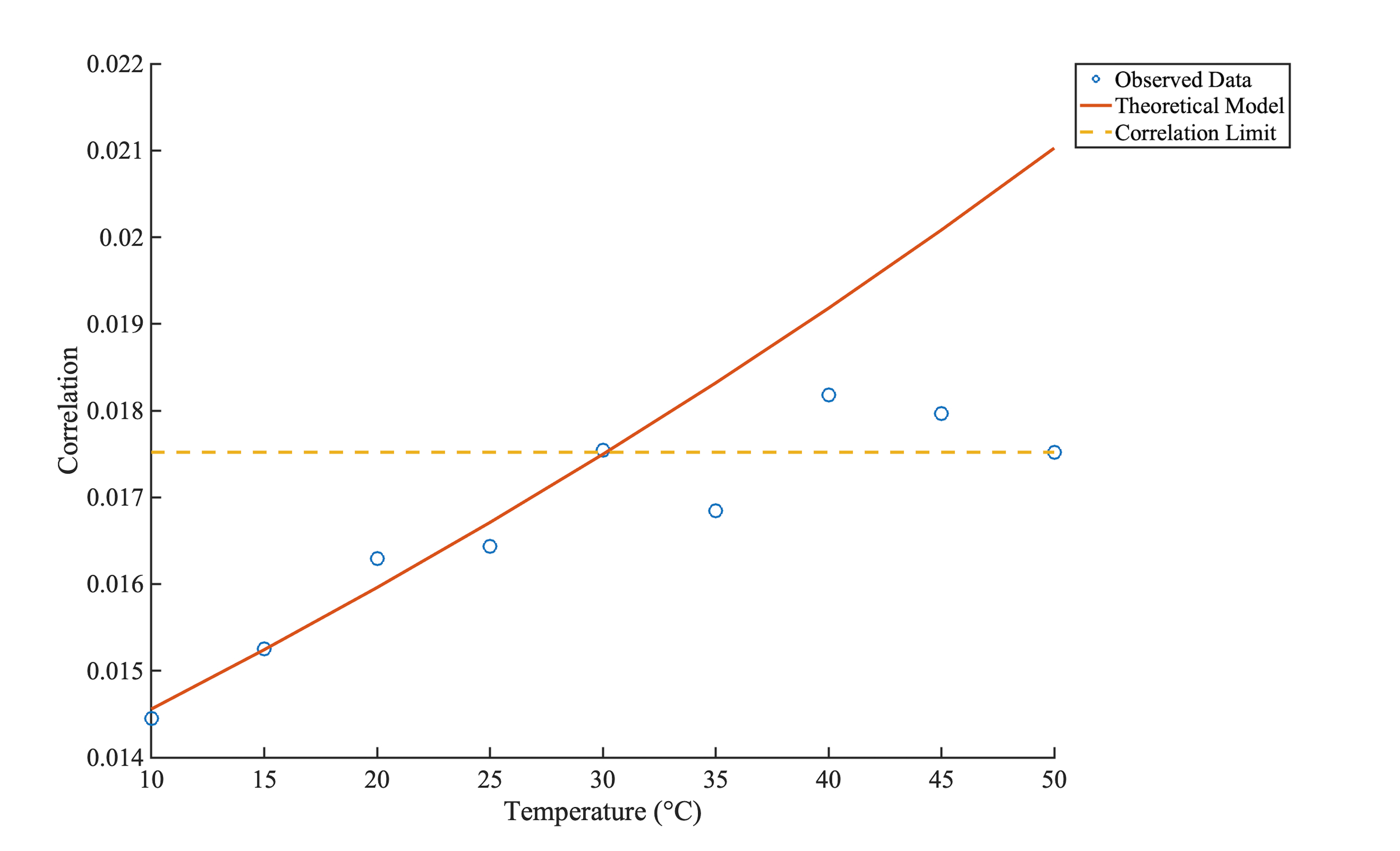}
\caption[]{Correlation verses temperature plot for camera three showing the correlation increase in accordance with the theoretical model to a limit which corresponds to the temperature of the image sets under test. Camera 3 Identified Temperature 30.15$^{\circ}$C}
\label{Camera06DSN}
\end{figure}

Using the theory presented in \cite{holst2007cmos} we apply a model based on the dark current density model seen in Equation \ref{DarkCurrentDensity} to the measured results. This model has the exponential form $y = a \mathrm{ e}^{bt}$. Each model resolves with an adjusted $R^2$ value of .9449, .9787 and .9523 respectively for camera 1, 2 and 3. These models are shown in Figure (\ref{AllCamerasDSN}) and then overlaid against the observed data in Figure \ref{Camera01DSN}, Figure \ref{Camera05DSN} and Figure \ref{Camera06DSN}. There is a strong indication that the correlation is related to the DSN as hypothesised. Taking the b value from each model we can approximate a value for the band gap energy $E_{t}-E_{G}$ for the cameras 1, 2 and 3 of 0.1896 eV, 0.3676 eV and 0.1268 eV. These values are consistent for band gap energies of silicon (1.1eV) and various dopant concentrations. Using this model we can identify that the correlation increases up to an approximately constant value. This constant value occurs when the temperature of the DSN reference pattern matches that of the image. Using this analysis, we can determine an approximate temperature for each image set. Camera 01 is identified as 30.5$^{\circ}$C, Camera 02 is identified as 28.35$^{\circ}$C and Camera 03 is identified as 30.15$^{\circ}$C.

It was observed that a prohibitively long time was required to take images at the exact temperature required for each set. As such, all images under test were taken over the range T=30$^{\circ}$C+/-2 due to the thermal balance of the equipment used. The temperature sensor used for these experiments was an MCP9808. The sensor bounced around the setpoint due to several reasons including observed self-warming on the sensor during image capture and environmental conditions. As such, at the extremes of temperature, it would become difficult to wait for the experimental apparatus to set upon a fixed temperature. To ensure a prohibitively long time was not needed to acquire images, all images were taken over a maximum 4$^{\circ}$C range of the target temperature. When averaged this could cause the expected temperature of the image set to be between 28$^{\circ}$C and 32$^{\circ}$C however, it is more likely than not that the average of the images would be 30$^{\circ}$C. Unfortunately, the temperature of the images under test was not recorded in the EXIF metadata meaning the exact temperature could not be determined. 

The MCP9808 temperature device used in these experiments has an overall accuracy of +/-0.25$^{\circ}$C. Accuracies should, therefore, add to obtain an overall error of +/-4.5$^{\circ}$C of the expected target temperature of 30$^{\circ}$C. Once this analysis is taken into account, we see that each camera has an accurate temperature result. We see that for each camera the forensic range should be as indicated in table \ref{TableTemps}. From here we see that the result from each camera is aligned to the actual range of the temperature taken during the image acquisition phase.

\begin{table} 
\centering
\caption{Identified Temperature of Image Sets} 
\begin{tabular}{cccc} 
	\hline
	& Identified Temperature ($^{\circ}$C) & Forensic Range ($^{\circ}$C)& Actual Range ($^{\circ}$C)  \\ 
	\hline
	Camera 01 & 30.5                   & 26.0 - 35.0    & 28.0 - 32.0   \\
	Camera 02 & 28.35                  & 23.85 - 32.85  & 28.0 - 32.0   \\
	Camera 03 & 30.15                  & 25.65 - 34.65  & 28.0 - 32.0   \\
	\hline
\end{tabular}
\label{TableTemps}
\end{table}



\section{Discussion}\label{discussion}

In our work presented here, it is not the first time DSN has been used as a forensic trace. However, it is the first time that a temperature dependency has been observed and used for forensic inference. While the work of \cite{kurosawa1999ccd,kurosawa2013casestudies} identified the use of DSN for unique forensic identification it is only now that we have further used the theory as presented in \cite{holst2007cmos} to determine the precise temperature that the image sensor was at during image capture. Determining temperature using SPN methods is contrary to the existing narrative seen in \cite{lukas2006digital,baabc} who suggest sensor pattern noise methods are immune to temperature variance. This is particularly relevant since in our work we have used DSN reference patterns against noise residuals which are traditionally used for PRNU based methods. Our image sets thus contain both DSN and PRNU making up PNU.

Critically in this work, we have demonstrated once again a bias present within the noise residues which is linked to the lens system of the camera. Observing lens effects in such a manner reinforces the work of \cite{matthews2018analysis, Matthews2018Isolation, Matthews2017Isolation} showing that the PRNU component of the PNU is excited differently due to the path discrete photons travel through each lens element. This is, however, contrary to the existing understanding of the SPN method whereby the image is tied uniquely to the image sensor with no contamination from sources which are not unique to the individual camera such as interchangeable lenses. Further investigation into this lens effect may lead to an understanding of why SPN based methods can provide discrimination of make and model and not just individual cameras as seen in \cite{akshatha2016digital}. Lenses may especially be relevant in \cite{akshatha2016digital} since mobile phone cameras are built with integrated lenses. 

Whereas the current literature has shown a distinct direction in optimising filters used for sensor-based methods \cite{lawgaly2017sensor} to create more precise results we have began a new focus to enable faster processing times. Such an approach may sacrifice precision. Lower precision does not lower the usefulness of the tool. The quicker exclusions can be made in the field means a focused effort can be made in a laboratory environment on evidence that requires robust examination. This would counter the current practice of bringing most devices back to digital forensic laboratories and allow a more efficient expenditure of resources. Excluding devices is particularly relevant in the application of images on what \cite{mislan2010growing} refer to as Type 2 and Type 3 mobile devices. These are devices in which large amounts of stored data including images which have the ability to store such data indefinitely. Efficient resourcing to allow time critical decisions is paramount when forensic intelligence is looked at in the military context where time can be of the essence due to the involvement of austere environments, sensitive and high profile targets and the resultant need for rapid execution \cite{Wilson2018}. By indicating how a simple, good enough ``Goldilocks'' filter can cut down on analysis times, we hope such tools can be adopted into on-site triage packages resulting in a more focused application of search warrants and more efficient use of forensic capabilities in the future. More work is required to verify the ability of such filters.

\section{Conclusion and Future Work} \label{conclusion}

In this paper, we have demonstrated a temperature bias in the method as first shown in \cite{lukas2006digital} and expanded upon in \cite{kurosawa2013casestudies}. This temperature bias present relates to the presence of dark current within an image and proves to be a useful forensic trace in its own right. We use this trace to isolate the temperature that an image is taken at independent of other sources such as EXIF metadata. This result is demonstrated across three CMOS image sensors of the same make and model and is experimentally linked to the dark current of the device.  Further, even when engineering designs are implemented to minimise the effects of dark current at the sensor silicon level, the effects of temperature on the SPN methods are still apparent indicating dark current is never completely removed. 

This study is, however, only an initial study into this physical phenomenon and should be conducted on a much broader sample. In the course of this study, we were exposed to a limitation in the ability of our lighting apparatus. This limitation resulted in an inability to vary the intensity of light that a scene was illuminated with. This work should be replicated using various exposure times and light intensity to measure the effects of dark current more thoroughly on the sensor pattern identification methods for the blind source camera problem. It is expected that light would also result in a variation in raw correlation obtained for each positive match. Such variances in correlation are expected to stack per the additive noise model and exacerbate the effects of either temperature or light intensity. 

Additionally, our small sample of image sensors means that this study is not an all-encompassing commentary on the issue of the thermal effect of SPN. Given the vast array of CCD sensors still in use, and the fact that their technology is significantly different to the active pixel CMOS sensors used in this study, verification on CCD sensors should be considered. Since the mobile imaging market is significantly increasing in size and complexity (for example Apple products with multiple sensors for a single image) a more extensive array of tests against CMOS sensors should also be conducted as future work. This initial study, however, has demonstrated a significant change in thinking to the current literature that SPN methods are immune to temperature dependency and thus, a new avenue of research is now opened to explore and exploit this phenomenon for forensic purposes. 


\section{Acknowledgements}
This research did not receive any specific grant from funding agencies in the public, commercial, or not-for-profit sectors. This work was supported with supercomputing resources provided by the Phoenix HPC service at the University of Adelaide. This research is supported by an Australian Government Research Training Program (RTP) Scholarship and forms part of a thesis chapter. The authors acknowledge the facilities, and the scientific and technical assistance, of the Australian Microscopy and Microanalysis Research Facility at Adelaide Microscopy, the University of Adelaide. This paper is part of a thesis.




  \bibliographystyle{elsarticle-num} 
  \bibliography{ICR2018-Temperature}

\end{document}